\documentclass[twocolumn,showpacs,prc,superscriptaddress]{revtex4}
\usepackage{amsmath,amssymb}
\providecommand{\LyX}{L\kern-.1667em\lower.25em\hbox{Y}\kern-.125emX\@}

\usepackage[T1]{fontenc}
\usepackage{graphics}
\usepackage{bm,hyphenat,xspace}
\makeatother
\newcommand{\mbck}{\bm{\mathcal{K}}}
\newcommand{\mck}{\mathcal{K}}
\begin{document}

\newcommand {\mbf}[1]{{\mathbf{#1}}}
\newcommand {\vecg}[1]{\mbox{\boldmath{$#1$}} }
\newcommand{\be}{\begin{eqnarray}}
\newcommand{\ee}{\end{eqnarray}}
\topmargin -1cm

\title{Spectroscopy of the unbound: One-neutron knockout reaction of $^{14}$Be}

\author{R.~ Crespo}
\affiliation{ Centro de F\'{\i}sica Nuclear, Universidade de Lisboa,
Av.\ Prof.\ Gama Pinto 2, 1649-003 Lisboa, Portugal}
\affiliation{ Departamento de F\'{\i}sica, Instituto Superior T\'ecnico, 
Taguspark, Av. Prof. Cavaco  Silva,  Taguspark, 2780-990 Porto Salvo, 
Oeiras, Portugal}
\author{A.~Deltuva}
\author{M.~ Rodr\'{\i}guez-Gallardo}
\author{E.~ Cravo}
\author{A.~C. Fonseca}
\affiliation{ Centro de F\'{\i}sica Nuclear, Universidade de Lisboa,
Av.\ Prof.\ Gama Pinto 2, 1649-003 Lisboa, Portugal}

%\email{raquel.crespo@tagus.ist.utl.pt//edgar@cii.fc.ul.pt
%//deltuva@cii.fc.ul.pt//mrodri@cii.fc.ul.pt//fonseca@cii.fc.ul.pt}

\date{\today}

\begin{abstract}
Full Faddeev-type calculations are performed for 
one-neutron knockout reaction of $^{14}$Be on proton target
at 69 MeV/u incident energy. 
Inclusive transverse momentum distributions for the 
outgoing ($^{12}$Be + n) system and 
semi-inclusive cross sections  are presented.
A significant proton-core single scattering contribution emerges where the valence neutron has non zero angular momentum relative to the core. This indicates
that distorted-wave impulse approximation is inadequate
and the complete multiple scattering series must be taken into
account for the considered reaction.
The magnitude of the semi-inclusive cross section
at quasifree scattering conditions is a clear signature of
the angular momentum of the valence nucleon.
\end{abstract}

\maketitle
%24.50.+g 	Direct reactions 
%25.40.Ep 	Inelastic proton scattering

\section{Introduction}

Knockout-reactions constitute a sensitive tool to investigate the
single-particle or cluster structure of nuclei. In the case of one-neutron knockout, the measurement
of the momentum distribution of the neutron and the A-1 system allows to identify, through the
shape of the distributions, the orbital angular momentum of the struck particle.

Recently  one-neutron knockout in the
$^{14}$Be scattering from a proton target at 69 MeV/u was measured at RIKEN
in order to obtain information on the unbound $^{13}$Be \cite{Kondo07} and the data analysed through the invariant mass method. This is a very powerful tool to study unbound nuclear 
states where  the invariant mass is determined by the momentum vectors of the 
outgoing particles \cite{Sugimoto}; the advantage of the
method is a very good energy resolution. Nevertheless,
very little is known about the $^{13}$Be system and contradictory results have
been found both from the theoretical and experimental sides.
Although it would be more appropriate to describe $^{14}$Be-p scattering in a 
4-body model ($^{12}$Be,n,n,p),
the exact 4-body scattering equations at present can only be 
solved for the four nucleon system below breakup threshold.
Therefore in this paper we use the 3-body 
Faddeev/Alt, Grassberger, and Sandhas (AGS) scattering 
formalism \cite{faddeev60,Alt,Glockle}. We explore  a number of $^{13}$Be
configurations and calculate the breakup observables
to shed light on the structure of the unbound system
and to stimulate further experimental work.

In the absence of exact many-body reaction calculations, 
Distorted Wave Impulse Approximation (DWIA) methods have been applied to
study such reactions \cite{Chant77}.
In this approach one  assumes that the incoming particle
collides with the struck nucleon as if it was free and the relative motion
of the particles in the entrance and exit channels
is described by distorted waves.
Moreover further approximations  are usually made in standard applications
of DWIA when evaluating the transition amplitude 
for the knockout scattering process $A(a,ab)B$ such as: 
(i) the potential approximation in the entrance channel
$V_{aA}-V_{ab} \sim V_{aB}(\mbf{r}_{aA})$;
(ii) the factorization approximation, which is only exact in 
Plane Wave Impulse Approximation  (PWIA); 
(iii) the on-shell approximation of the transition amplitude. 
Although the validity of some of these approximations has not been properly investigated yet, it has been shown \cite{Crespo08} that DWIA leads to an incomplete and truncated multiple scattering expansion  that is responsible for inaccurate results of $^{11}$Be-p breakup observables at intermediate energies. The work in Refs.~\cite{Crespo08,Crespo07b,Del07}  is based on exact 
Faddeev/AGS theory \cite{faddeev60,Alt,Glockle} and provides not only a benchmark to study approximate methods commonly used in nuclear reactions, but also a means to obtain a better description of the dynamics that drives knockout reactions involving light bound systems.

\section{The Faddeev/AGS equations}

Let us consider 3 different particles denoted 1,2,3,
interacting by means of two-body potentials.
The Faddeev/AGS multiple scattering framework is a three-body scattering
formalism that treats all open channels 
(elastic, inelastic, transfer and breakup) on equal footing. 
It is therefore adequate to describe the scattering of light bound systems 
such as  halo nuclei
where breakup thresholds are close to the ground state.
In the Faddeev/AGS framework the full scattering amplitude may be expressed 
as a multiple scattering expansion in terms 
of the two-body transition operators for each interacting pair.
We summarize here the main expressions of the formalism, using the 
odd man out notation appropriate for 3-body problems
which means, for example, that the interaction between
the pair (23) is denoted as $v_1$. Assuming that the system is 
non-relativistic, one writes the total Hamiltonian as
\be
H = H_0 + \sum_\gamma v_\gamma \; ,
\ee
with the kinetic energy operator $H_0$ and
the interaction $v_\gamma$ for the pair $\gamma$.
The Hamiltonian can be rewritten as
\be
H = H_\alpha + V^\alpha \; ,
\ee
where $H_\alpha$ is the Hamiltonian for channel $\alpha$
\be
H_\alpha = H_0 + v_\alpha \; ,
\ee
and $V^\alpha$ represents the sum of interactions external to partition $\alpha$
\be
 V^\alpha = \sum_{\gamma \neq \alpha} v_\gamma \; .
\label{Valpha}
\ee
The $\alpha = 0$ partition corresponds to three free particles in the 
continuum where $V^0$ is the sum of all pair interactions. 
Let us consider the scattering from the 
initial state $\alpha$ to the final state $\beta$. 
The operators $U^{\beta \alpha}$, whose on-shell matrix elements are the transition amplitudes,  are obtained by solving the three-body AGS  integral equation~\cite{Alt,Glockle} that reads
\be
U^{\beta \alpha} = \bar{\delta}_{\beta \alpha} G_0^{-1}
+ \sum_{\gamma} U^{\beta \gamma}G_0 t_\gamma \bar{\delta}_{\gamma \alpha} \; ,
\ee
or
\be
U^{\beta \alpha} = \bar{\delta}_{\beta \alpha} G_0^{-1}
+ \sum_{\gamma} \bar{\delta}_{\beta \gamma }
t_\gamma G_0 U^{\gamma \alpha} \; ,
\label{Uba2}
\ee
where $ \bar{\delta}_{\beta \alpha} = 1 - {\delta}_{\beta \alpha}$.
The pair transition operator is
\be
 t_{\gamma} =  v_{\gamma} +  v_{\gamma} G_0  t_{\gamma} \; ,
\ee 
where $ G_0 $ is the free resolvent
\be
G_0 = (E+i0 - H_0)^{-1},
\ee
and $E$ the total energy of the three-particle system in the center of mass (c.m.) frame.
For breakup ($\beta = 0$ in the final state) one has 
\be
U^{0 \alpha} = G^{-1}_0 + \sum_\gamma t_\gamma G_0 U^{\gamma \alpha}, 
\label{U0alpha}
\ee
where $U^{\gamma \alpha}$  is obtained from the solution of Eq.~(\ref{Uba2}) with 
$\alpha, \beta, \gamma = (1,2,3)$.  
The scattering amplitudes are the matrix elements of 
$U^{\beta \alpha}$  calculated between initial and final states that are
eigenstates of the corresponding channel Hamiltonian 
$H_\alpha \, (H_\beta)$ with the same energy eigenvalue $E$. For breakup the final state is the product of two plane waves corresponding 
to the relative motion of three free particles that may be expressed in 
any of the relative Jacobi variables. In the latter case the contribution 
of the  $G_0^{-1}$ term is zero.

\begin{figure}{\par\centering \resizebox*{0.1\textwidth}{!}
{\includegraphics{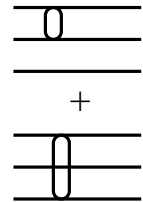}} \par}\caption{\label{Fig:single} 
Single scattering diagrams for breakup in the Faddeev/AGS scattering framework.}
\end{figure}

The solution of the Faddeev/AGS equations can be found by iteration leading to
\be
U^{0 \alpha} &=& \sum_\gamma t_\gamma  
\bar{\delta}_{ \gamma  \alpha } + \sum_\gamma t_\gamma 
\sum_\xi G_0 \bar{\delta}_{\gamma \xi }  t_\xi
\bar{\delta}_{\xi  \alpha} \nonumber \\ &+&
\sum_\gamma t_\gamma 
\sum_\xi G_0  \bar{\delta}_{\gamma \xi }  t_\xi
\sum_\eta G_0 \bar{\delta}_{ \xi \eta} t_\eta  \bar{\delta}_{\eta \alpha}
 \nonumber \\ &+& \cdots \; ,
\label{AGS}
\ee
where the series is summed up by the Pad\'e method \cite{Pade}.
The successive terms of this series can be considered as first
order (single scattering), second order (double scattering) and so
on in the transition operators. For the breakup process where particle 1 scatters from pair (23) into the breakup channel (1,2,3), the breakup series in single scattering 
is represented diagrammatically in Fig.~\ref{Fig:single}
where the upper particle is taken as particle 1.

In our calculations Eqs.~(\ref{Uba2}-\ref{AGS}) are solved exactly 
in momentum space after partial wave decomposition 
and discretization of all momentum variables. 
We include the nuclear interaction between all three pairs,
and the Coulomb interaction between the proton and $^{13}$Be, 
following  the technical developments implemented in 
Refs.~\cite{Del05b,Del06b} for proton-deuteron and $\alpha$-deuteron
elastic scattering and breakup that were also used 
in Refs.~\cite{Crespo07b,Del07} to study  p-$^{11}$Be elastic scattering and
breakup. 

%%%%%%%%%%%%%%%%% single breakup
%%%%%%%%%%%%%%%%%%%%%%%%%%%%%%%%%%%%%%%%%%%%%%%%%%%%%%%%%%%%%%

\section{The physical content of the single scattering term at np QFS}

Let us consider a reaction where the proton (particle  1) scatters from a
bound pair of particles (23) into the breakup channel $(1,2,3)$; 
particle 2 is a valence neutron and particle 3 the $^{13}$Be core. 
Actual experiments involving  
halo nuclei  are performed in inverse kinematics with the
scattering of a radioactive halo beam from a stable target. The results
discussed here are however independent of the kinematics that is used.

We have shown \cite{Crespo08} that at sufficiently high energies
and for some suitable kinematics configurations
the single scattering term  provides a reasonable
approximation to the multiple scattering series. 
In addition, the single scattering component where the projectile strikes 
the valence particle (first diagram in Fig.~\ref{Fig:single})
is dominant in the case of p($^{11}$Be,$^{10}$Be n)p
\cite{Crespo08}. Therefore, we analyze here in detail the physical content of the two components of the 
single scattering term involving both the scattering from the struck
neutron and the $^{13}$Be core.

In this section we shall use capital letters for momenta in the LAB frame
and lower case letters for the c.m. frame.

In the case of the single scattering from the struck neutron, the transition amplitude from an initial state 
$ | \psi_1 \rangle = |\mbf{k}_1 \phi_{23} \rangle $ to a final state
$ | \mbf{k}^{\prime}_1 \mbf{k}^{\prime}_2   \mbf{k}^{\prime}_3 \rangle$
 is given by the matrix elements of the transition operator   $t_3= t_{12}(\omega_{12})$ with relative pair (1,2) energy $\omega_{12}$.  As usual $ \mbf{k}_1$ is the initial relative momentum between particle 1 and the c.m. of pair (23). In the final state, $ \mbf{k}^{\prime}_1,  \mbf{k}^{\prime}_2$, and $ \mbf{k}^{\prime}_3$ are the final momenta for the three free particles in the three-body c.m. frame where $ \mbf{k}^{\prime}_1+  \mbf{k}^{\prime}_2  +  \mbf{k}^{\prime}_3 = 0$.

The breakup transition amplitude, corresponding to the single scattering approximation (SSA),
 may be written as 
\be
\langle \mbf{k}^{\prime}_1 \mbf{k}^{\prime}_2 \mbf{k}^{\prime}_3 |t_{3}| \psi_1 \rangle = \langle \mbf{q}^{\prime}_{12} |t_{12} (\omega_{12})| \mbf{q}_{12}\rangle \phi_{23} (\mbf{q}_{23})~,
\label{3kst12}
\ee
where
\be
\mbf{q}^{\prime}_{12} = \mbf{k}^{\prime}_1 + \frac{m_1}{M_{12}} \mbf{k}^{\prime}_3 ~,
\label{qÌð12}
\ee
\be
\mbf{q}_{12} = \mbf{k}_1 + \frac{m_1}{M_{12}} \mbf{k}^{\prime}_3 ~,
\label{q12}
\ee
\be
\mbf{q}_{23} = - \mbf{k}^{\prime}_3 - \frac{m_3}{M_{23}} \mbf{k}_1 ~,
\label{q23}
\ee
\be
\omega_{12} = E - \frac{k^{\prime ^2}_3}{2 \mu_{3(12)}} = \frac{q^{\prime ^2}_{12}}{2 \mu_{12}} ~,
\label{omega12}
\ee
with $M_{ij} = m_i + m_j$, $\mu_{ij} = \frac{m_i m_j}{M_{ij}}$, $\mu_{i(jk)} = \frac{m_i M_{jk}}{M}$ and $M = m_i + m_j + m_k$. Therefore the two-body t-matrix $t_{12}$ is calculated at half the energy shell.

At the np quasifree scattering (QFS) kinematical condition  particle 3  (the $^{13}$Be core) acts as a spectator which means that 
in the Lab frame ${\mbf{K}}_3 = 0$ corresponding to $\mbf{k}^{\prime}_3 = - \frac{m_3}{M_{23}} \mbf{k}_1$ in the c.m. frame. Therefore, according to Eq. (\ref{q23}) 
\be
[\mbf{q}_{23}]^{{\rm QFS}} = 0 ~,
\label{q23=0}
\ee
and
\be
[\omega_{12}]^{{\rm QFS}} = E \frac{M \; m_2}{M_{23}\; M_{12}} = \frac{[q^{2}_{12}]^{{\rm QFS}}}{2 \mu_{12}} 
\label{+omega12}
\ee
in the limit of zero binding energy for the pair 23 as demonstrated in the Appendix A. Only in such zero binding limit and in the QFS kinematical condition does the t-matrix $t_{12}(\omega_{12})$ gets calculated on the energy shell. Away from the QFS kinematical condition the $t_{12}(\omega_{12})$ matrix elements become half-on-shell
and the total transition amplitude probes the non zero relative momentum
components of the bound pair.
Higher order multiple scattering terms necessarily probe
off-energy-shell effects and higher relative momentum components, even at the QFS conditions.

We now consider the SSA component where the proton scatters from the core.
The transition amplitude is then given by
\be
\langle \mbf{k}^{\prime}_1 \mbf{k}^{\prime}_2 \mbf{k}^{\prime}_3 |t_{2}| \psi_1 \rangle = \langle \mbf{q}^{\prime}_{31} |t_{13} (\omega_{13})| \mbf{q}_{31}\rangle \phi_{23} (\mbf{q}_{23}),
\label{3kst13}
\ee
with $\mbf{q}^{\prime}_{31}  =  - \frac{m_1}{M_{13}} \mbf{k}^{\prime}_2  - \mbf{k}^{\prime}_1$, $\mbf{q}_{31}  =  - \frac{m_1}{M_{13}} \mbf{k}^{\prime}_2  - \mbf{k}_1$, \linebreak $\mbf{q}_{23}  =  \mbf{k}^{\prime}_2 + \frac{m_2}{M_{23}} \mbf{k}_1$,  and $\omega_{13} = E - \frac{k^{\prime ^2}_2}{2 \mu_{2(13)}}$. 

In the np QFS limit where $\mbf{k}^{\prime}_3 = -\mbf{k}^{\prime}_1 - \mbf{k}^{\prime}_2 = - \frac{m_3}{M_{23}} \mbf{k}_1$ we get
\be
\mbf{k}^{\prime}_2 = - \mbf{k}^{\prime}_1 + \frac{m_3}{M_{23}} \mbf{k}_1 ~,
\label{kÌð2}
\ee
leading to
$$
[\mbf{q}_{23}]^{{\rm QFS}} = \mbf{k}_1 - \mbf{k}^{\prime}_1 = \mbf{\bar{k}} ~,
$$
where $\mbf{  \bar{k} }$ is the momentum transfer of particle 1, the projectile. Therefore, in the case of the projectile scattering from the core, the wave function of the target nucleus is probed at nonzero relative momentum.
For the case of the reaction  p($^{11}$Be,$^{10}$Be n)p studied in 
\cite{Crespo08}, the valence neutron is bound to $^{10}$Be in S-wave whose wave function momentum  distribution is sharply peaked around zero momentum. Thus,  in this case, the wave function, when probed at larger
momentum $q_{23}$, is already very small, leading to a scattering contribution from the core that is relatively small. For other orbital angular momentum states such as P- or D-waves, the wave function
becomes nonnegligible at larger $q_{23}$ and the contribution 
from projectile-core
scattering may become significant. Therefore
standard DWIA calculations which only take into account the contribution
from the scattering of the struck particle may become inadequate in the
case of struck particles bound with nonzero orbital angular momentum.
We shall return to this point later.

\section{The breakup observables in inverse kinematics}

We now consider the breakup of a
radioactive beam involving a two-body halo nucleus assumed to be well described 
by a core and a valence neutron; the halo nucleus collides with a proton 
target leading to three free particles in the final state.
This final state is described in terms of 9 kinematical variables. Momentum and
energy conservation reduces this number to 5 independent variables.
With the recent developments at  the radioactive beam facilities 
it is now possible to measure fivefold fully exclusive observables. By further
integration upon the variables of the emitted particles 
semi-inclusive as well as inclusive breakup observables can be measured 
although with loss of physical information.

In actual experiments performed in inverse kinematics it is the
halo core which is measured 
either directly or by reconstruction from the other emitted fragments.
We therefore chose the Jacobi momenta 
\be
\mbf{p} &=& \frac{m_n \mbf{K}_p - m_p \mbf{K}_n}{m_p + m_n} ~,
\nonumber \\
\mbf{q} &=& \frac{ (m_p+m_n)\mbf{K}_C - m_C (\mbf{K}_p +\mbf{K}_n )}{M} ~,
\label{pqjacobian}
\ee
with $\mbf{K}_p$, $ \mbf{K}_n$, $ \mbf{K}_C$ ($m_p, m_n, m_C$) being the LAB momenta (masses)
of the proton, valence
neutron and core in the exit system, and $M= m_p + m_n + m_C$.
The breakup differential cross section is calculated from the 
on-shell matrix elements of the AGS operators, 
$T^{0\alpha}=\langle \mbf{q} \mbf{p}| U^{0\alpha} | \psi_\alpha \rangle$
where particle $\alpha$ is the spectator in the initial state 
(in our case $\alpha$ is the proton) and 
the Jacobi momenta in the final state satisfy the on-shell relations
\be
E - \frac{p^2}{2 \mu} - \frac{q^2}{2  \overline{\mu} } = 0 ~,
\ee
with the reduced masses
\be
\mu &=& \frac{m_n m_p}{m_n+m_p} ~,\nonumber \\
 \overline{\mu}  &=&  \frac{m_C(m_n + m_p)}{M} ~.
\ee

\subsection*{Fully exclusive observables}

\begin{figure}
{\par\centering \resizebox*{0.45\textwidth}{!}
{\includegraphics{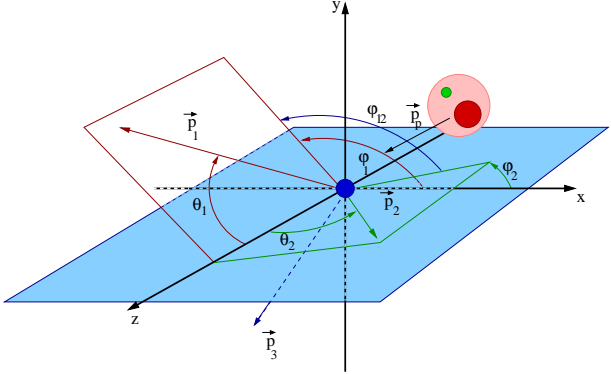}} \par}
\caption{\label{Fig:kinem}
 (Color online) Kinematic angles for breakup in inverse kinematics in the LAB
frame
}
\end{figure}

The fully exclusive breakup observables are measured in the LAB system.
The  kinematic configuration of three-body breakup is characterized 
by the polar and  azimuthal angles $\Omega_i=(\theta_i, \phi_i)$ 
of the two detected particles as in  Fig.~\ref{Fig:kinem} which we assume
 to be the core C and the valence neutron n.

In order to calculate this observable we begin by writing 
the general expression in terms of the on-shell matrix elements of
the AGS operators an including
momentum and energy conservation
\begin{widetext}
\be
\frac{d^5\sigma}{d\mbf{\hat{K}}_n d \mbf{K}_C} 
 &=& (2 \pi)^4 \frac{m_n + m_C}{K_{\rm LAB}} 
\int d \mbf{K}_p K_n^2 dK_n
|T^{0\alpha}|^2 
 \delta( \mbf{K}_{\rm LAB} - \mbf{K}_n - \mbf{K}_p - \mbf{K}_C)
\delta(E_{\rm LAB} + \epsilon - \frac{K_p^2}{2 m_p} 
- \frac{K_n^2}{2 m_n}
- \frac{K_C^2}{2 m_C}
 ) 
\nonumber \\
&=& (2 \pi)^4  \frac{m_n + m_C}{K_{\rm LAB}} 
\int  K_n^2 dK_n
|T^{0\alpha}|^2 
\delta(E_{\rm LAB} + \epsilon - 
\frac{(\mbf{K}_{\rm LAB} - \mbf{K}_n - \mbf{K}_C)^2}{2 m_p} 
- \frac{K_n^2}{2 m_n}
- \frac{K_C^2}{2 m_C}
 ) 
\nonumber \\
&=&  (2 \pi)^4  \frac{m_n + m_C}{K_{\rm LAB}} 
 m_p m_n \sum_i \left[ |T^{0\alpha}|^2
\frac{  K_n^2   }    
{ | (m_n+m_p)K_n -  m_n ( \mbf{K}_{\rm LAB} - \mbf{K}_{C}) \cdot \mbf{\hat{K}}_n| }
\right]_i ~,
\label{fully1}
\ee
\end{widetext}
where the sum on $i$ involves the momenta $K_n$ given by the zero's of the
argument of the energy conserving $\delta$-function
\be
E_{\rm LAB} + \epsilon - 
\frac{(\mbf{K}_{\rm LAB} - \mbf{K}_n - \mbf{K}_C)^2}{2 m_p} 
- \frac{K_n^2}{2 m_n}
- \frac{K_C^2}{2 m_C} = 0~, 
\nonumber \\
\label{zerodelta-energy}
\ee
$\mbf{K}_{\rm LAB}$ is the beam momentum in the Lab frame and $E_{\rm LAB}$ the corresponding energy. To arrive to Eq.~(\ref{fully1}) we used the property of  the $\delta$-function, 
\be
\delta(g(x))= \sum_i 
\frac{\delta(x-x_i)}{ |g^\prime(x_i)|  } ,
\label{deltaproperty}	
\ee
where the $x_i$ s are the zeros of the function $g(x)$.
Eq.~(\ref{zerodelta-energy}) defines an ellipse in the $K_n-K_C$ plane. 
The points lying on that ellipse or on the corresponding curve in the $E_n-E_C$ 
plane comprise the kinematically allowed S-curve on 
which the physically accessible events have to lie \cite{Glockle}.
The phase space factor in Eq.~(\ref{fully1}) has the disadvantage that
it diverges at the $K_C$ values at which the ellipse has an infinite derivative.
As a result one often replaces the dependence on the energy $E_C$ by 
the arclength $S$  related to the LAB energies $E_n$ and  $E_C$ of the 
two detected particles as  $S = \int_0^S dS$  \cite{chmielewski:03a} with
\be
dS &=& \sqrt{dE_n^2 + dE_C^2} = 
 dE_C  \sqrt{1 + \left(\frac{dE_n}{dE_C}\right)^2}
\nonumber \\
 &=& dE_C  \sqrt{1 + \left(  \frac{m_C K_n}{m_n K_C}      
\frac{dK_n}{dK_C}\right)^2}~~.
\ee
By differentiating the argument of the energy conserving $\delta$-function with respect to $K_C$ and
taking $K_n = K_n(K_C)$ we obtain
\be
  \frac{dK_n}{dK_C}   =  -
\frac{(m_p + m_C)K_C - m_C(\mbf{K}_{\rm LAB}-\mbf{K}_n)\cdot
\mbf{\hat{K}}_C }
{ (m_p + m_n)K_n - m_n(\mbf{K}_{\rm LAB}-\mbf{K}_C)\cdot \mbf{\hat{K}}_n    }
 \frac{m_n}{m_C} ~. 
\label{dkn}
\nonumber
\\
\ee
Therefore one can measure the fully exclusive 
fivefold differential breakup cross section 
$ {d^5 \sigma}/{d\Omega_n d\Omega_C dS}$ where, from Eq.~(\ref{fully1}), we get 
\begin{widetext}
\be
\frac{d^5\sigma}{d\Omega_n d\Omega_C   dS} 
 = (2 \pi)^4  \frac{m_n + m_C}{K_{\rm LAB}}  
m_p m_n m_C K_C  \left\{ |T^{0\alpha}|^2
\frac{  K_n^2   }    
{ | (m_n + m_p)K_n -  m_n( \mbf{K}_{\rm LAB} - \mbf{K}_{C}) \cdot \mbf{\hat{K}}_n |} \right. \nonumber \\ 
\times \left. \left[1 + \left(  \frac{m_C K_n}{m_n K_C}      
\frac{dK_n}{dK_C}\right)^2\right]^{- \frac{1}{2}}
\right\}_i~,
\label{fully}
\ee
\end{widetext}
with $dK_n/dK_C$ given by Eq.~(\ref{dkn}). The sum on $i$ disappears in Eq.~(\ref{fully}) relative to Eq.~(\ref{fully1}), because there is a one to one correspondence between $E_n$ and $S$ which may not exist between $E_n$ and $E_C$.

%%%%%%%%%%%%%%%%%%%%%%%%%%%%%%%%%%%%%%%%%%%%
%%%%%%%%%%%%%%%%%%%%%%%%%%%%%%%%%%%%%%%%%%%%

\subsection*{Semi-inclusive observables}

The semi-exclusive differential cross section 
$ {d^3 \sigma}/{d\Omega_C  dE_C}$ may be obtained from
the fivefold differential cross section \eqref{fully1}
by integrating over the angles of the emitted valence neutron.
However, it is more convenient to start from 
the fivefold differential cross section in the c.m. frame
\be
\frac{d^5\sigma}{d^2\hat p \;d^3q}  \!\!=\!\! (2 \pi)^4  \frac{m_n \!+\! m_C}{K_{\rm LAB}}
\! \!\!\int\! \!|T^{0\alpha}|^2 
 \delta(E \!- \!\frac{p^2}{2 \mu} \!-\!  \frac{q^2}{2 {\overline  \mu}}) 
p^2 dp.
\ee
Using the property~(\ref{deltaproperty}) of  the $\delta$-function one gets
\be
\frac{d^5\sigma}{d^2\hat p \;d^3q}  = (2 \pi)^4  \frac{m_n \!+\! m_C}{K_{\rm LAB}} |T^{0\alpha}|^2 
\mu \sqrt{2 \mu E - \frac{\mu}{\overline\mu}q^2}~.
\ee
From Eq.~(\ref{pqjacobian}), using $\mbck_{\rm TOT}=\mbck_n + 
\mbck_p + \mbck_C$ and remembering that in the chosen Jacobi
set $\mbf{q} =\mbck_C - \frac{m_C}{M}\mbck_{\rm TOT}$ one gets
\begin{widetext}
\be
\frac{d^3\sigma}{d\Omega_C  dE_C   }  = (2 \pi)^4 
\frac{m_n + m_C}{K_{\rm LAB}} m_C \mck_C
\int d^2\hat p |T^{0\alpha}|^2 
\mu \sqrt{2 \mu E - \frac{\mu}{\overline\mu}
\left(\mck_C^2 + \frac{m_C^2}{M^2}\mck_{\rm TOT}^2 - 2 \frac{m_C}{M}
\mbck_C \cdot \mbck_{\rm TOT}  \right)
} ~,
\ee
\end{widetext}
where the calligraphic momenta may be LAB or c.m. momenta depending on $\mbck_{\rm TOT} = K_{\rm LAB}$ or  $\mbck_{\rm TOT} = 0$. 
Clearly in  the integrand of this equation the square root must be positive
definite and, therefore, when the LAB energy of the emitted core 
exceeds a certain limit, $E_C^{\rm LAB}$(max), which depends on the 
angle of the emitted core, then the semi-exclusive
cross section should vanish. In order to find the rate at which the cross
section vanishes one should take the derivative of the differential cross 
section with respect to the energy of the core. It follows that 
at $E_C^{\rm LAB}$(max)  this derivative is infinite and therefore the 
semi-inclusive cross section should exhibit almost a sharp cutoff at the 
maximum value of the core laboratory energy unless the breakup amplitude
is very small close to the maximum energy.
Hence the energy behavior of the semi-inclusive cross section
at high energies follows from a delicate interplay between the phase
space and the scattering amplitude which should thus be accurately calculated.

\subsection*{Inclusive observables}

As shown in the Appendix B, from this semi-inclusive cross section one 
can calculate the inclusive perpendicular momentum distribution for one of
the detected particles, say the core, as
\be
\frac{d \sigma}{d K_C^p} 
= 2 \pi K_C^p  \int_{-\infty}^{+\infty}\frac{d^3 \sigma}{d^3 K_C}dK_C^z
\ee
and the inclusive transverse momentum distribution as 
\be
\frac{d \sigma}{d K_C^x} 
= 2   \int_{-\infty}^{+\infty}\int_{0}^{+\infty} 
\frac{d^3 \sigma}{d^3 K_C} dK_C^z  dK_C^y
\ee
 with 
\be
\frac{d^3 \sigma}{d^3 K_C}= 
\frac{1}{ m_C K_C} \frac{d^3 \sigma}{ d\Omega_C dE_C }~.
\ee

\section{The pair interactions}\label{The pair interactions}

In this work we address the scattering of the ($^{13}$Be +n) system
by a proton target. 
Before solving Faddeev/AGS equations we need to define
each pair interaction, that is, the  p-n, p-$^{13}$Be and n-$^{13}$Be
interactions.

For the p-n we take the realistic nucleon-nucleon CD Bonn potential 
\cite{CDBONN}.

For the potential between 
the proton and $^{13}$Be core we use a phenomenological
optical model with parameters taken from the Watson global
optical potential parametrization \cite{Watson69,Crespo07b}.
The energy dependent parameters of the optical potential are taken 
at the proton laboratory energy of the p-$^{14}$Be reaction in direct 
kinematics.

The interaction between the valence neutron and the $^{13}$Be core 
in $^{14}$Be   depends on the orbital angular configuration 
of the valence nucleon for 
the $^{13}$Be + n cluster system. Since the total angular momentum of the core
and valence neutron are coupled to $^{14}$Be($0^+$), it means that the 
the total angular momentum of the valence nucleon is identical to the $^{13}$Be core.
Little is known about this system with exception of a possible
$5/2^+$ (D-wave) resonance with a relative energy 
$E_{\rm rel}$($^{12}$Be+n)$ \sim $2 MeV above the neutron breakup threshold.    

In this work we take the spectroscopic study of $^{13}$Be obtained from
 the proton-induced reaction on $^{14}$Be at 69 MeV/u in inverse kinematics
performed by the invariant mass method at RIKEN \cite{Kondo07}.
The  $^{13}$Be   resonances of 
$1/2^-$ (P-wave) with  $E_{\rm rel}=0.45$ MeV, 
$1/2^+$ (S-wave) with  $E_{\rm rel}=1.17$ MeV, and
$5/2^+$ (D-wave) with  $E_{\rm rel}=2.34$ MeV
were assigned through the analysis of the transverse momentum distributions
of the outgoing $^{13}$Be  system. Following this work we 
take three possible single-particle configurations for $^{14}$Be(0$^+$)
\be
|^{14}{\rm Be} \rangle = |^{13}{\rm Be} (1/2^-)\otimes {\rm n}(1p1/2)\rangle, \\
|^{14}{\rm Be}  \rangle = |^{13}{\rm Be} (1/2^+)\otimes {\rm n}(2s1/2)\rangle, \\ 
|^{14}{\rm Be}  \rangle = |^{13}{\rm Be} (5/2^+)\otimes {\rm n}(1d5/2)\rangle,
\ee
with binding energy $\epsilon = E_{\rm rel} + S_{2n}$,  $S_{2n}=1.26$ MeV
being the two neutron separation energy of $^{14}$Be.

\begin{figure}
{\par\centering \resizebox*{0.40\textwidth}{!}
{\includegraphics{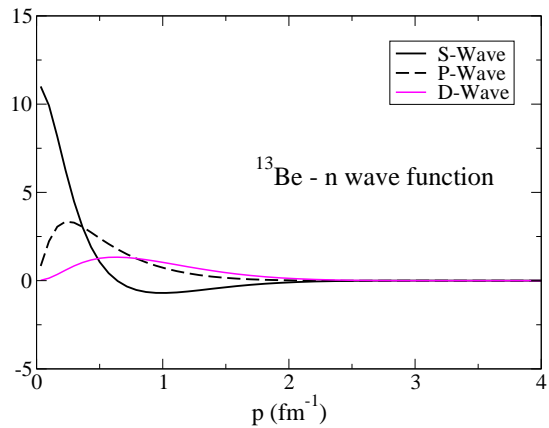}} \par}
\caption{\label{Fig:wf}
 (Color online) Relative wave function of the valence neutron and $^{13}$Be core in momentum space for S-(dark solid  line),
P-(dark dashed line) and D-wave bound states (light solid line)
}
\end{figure}

The interaction between the valence neutron and the $^{13}$Be core 
is taken to be of the form
\be
V(r)= - V_c f(r,R_0,a_0) 
\ee
where $f(r,R,a)$ is the usual Woods-Saxon form factor
\be
f(r,R,a) = 1/\{1 + \exp [(r-R)/a] \},
\ee
and $R_i$ = $r_i A^{\frac{1}{3}}$. 
The potential parameters for each single-particle
configuration are listed in Table \ref{Tab:n-13Be} and the corresponding wave functions shown in Fig.~\ref{Fig:wf}.

\begin{table}[!ht]
\caption{\label{Tab:n-13Be} Parameters of the n-$^{13}$Be interaction plus 
$r_0= $ 1.2 fm and $a_0=$ 0.6 fm.}
\begin{center}\begin{tabular}{|c|c|}
\hline
   State & $V_c$(MeV) 
\tabularnewline
\hline
    1p1/2          &   32.922    
\tabularnewline
    2s1/2          &   61.667 
 \tabularnewline
    1d5/2          &   68.980  
 \tabularnewline
\hline
\end{tabular}\end{center}
\end{table}

In order to access the contribution of other partial waves in the n-$^{13}$Be interaction beyond the one that is responsible for the single particle configuration leading to the $0^+$ ground state of $^{14}$Be, we take in those partial waves the potential corresponding to $V_c=32.922$ MeV in  Table~\ref{Tab:n-13Be}. Therefore $0^+$  partial waves are driven by the potentials in Table~\ref{Tab:n-13Be}, depending on the choice of  $^{13}$Be resonance we consider; all other partial waves are driven by the weakest potential of all three for no other reason than lack of a more enlightened choice.

%%%%%%%%%%%%%%%%%%%%%%%%%%%%%%%%%%%%%%%%%%%%
%%%%%%%%%%%%%%%%%%%%%%%%%%%%%%%%%%%%%%%%%%%%

\section{Results}

In the solution of the Faddeev/AGS equations we include n-p partial waves with relative orbital angular momentum $\ell_{np} \leq 8$, and p-$^{13}$Be with   $L \leq 10 $.
For the  n-$^{13}$Be  we start by including just the $0^+$ partial wave using the interactions in Table~ \ref{Tab:n-13Be} depending on the specific choice of spin and parity we consider for $^{13}$Be, that is, $1/2^-$, $1/2^+$, or $5/2^+$. Three-body total angular momentum is included up to 100 and in all calculations  unphysical bound states are properly removed \cite{Del07} keeping
a single bound state of $^{14}$Be in  $0^+$ partial wave.

We present in this section the calculated semi-inclusive breakup observables around QFS and the inclusive momentum distributions of the $^{13}$Be system.

In inverse kinematics the exact np QFS kinematic conditions 
correspond to a $^{13}$Be LAB momentum of ${\bf K}_{C}= \frac{m_{C}}{M_{Cn}}{\bf K}_{Cn} $, that is, taking $^{14}$Be in the ZZ direction we have $( E_{C}^{\rm LAB},\theta_{C}^{\rm LAB},\phi_{C}^{\rm LAB})^{\rm QFS} = (\frac{m_{C}}{M_{Cn}}E_{Cn},0,0)$, where the index $C$ denotes $^{13}$Be, $Cn$ denotes $^{14}$Be and $M_{Cn} = m_C + m_n$.  For $^{14}$Be scattering from a proton at 69 MeV/u this corresponds  to  $(897 ~{\rm MeV},0,0)$.

\begin{figure*}
{\par\centering \resizebox*{0.60\textwidth}{!}
{\includegraphics{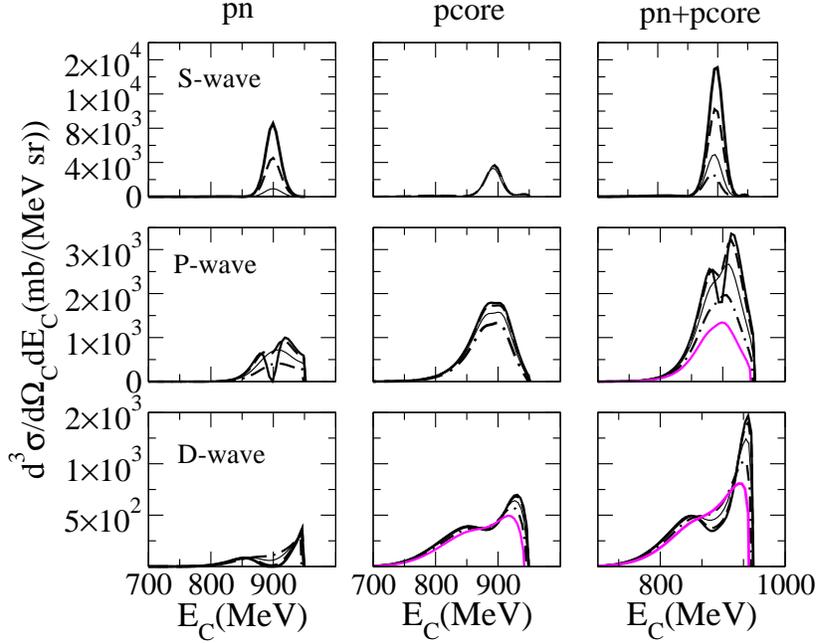}} \par}
\caption{\label{Fig:ds-e69-ssca-spd}
 (Color online) Semi-inclusive cross section for the breakup
  p($^{14}$Be,$^{13}$Be)np at 69~MeV/u  using the
SSA contribution for $\theta_{C}=0^\circ$ (dark thick solid line), 
$\theta_{C}=0.4^\circ$ (dashed line),
$\theta_{C}=0.8^\circ$ (dark thin solid line),
$\theta_{C}=1.2^\circ$ (dashed dotted line), and
$\theta_{C}=1.4^\circ$ (light solid line).
}
\end{figure*}

In Fig.~\ref{Fig:ds-e69-ssca-spd} we show the semi-inclusive cross section for the breakup
p($^{14}$Be,$^{13}$Be)np at 69~MeV/u using the SSA. We separately consider
the scattering from the valence neutron, the scattering from the core, and the scattering from both the valence and the core. As shown in Sec. III, the cross section calculated using SSA 
from the struck particle probes the bound state wave function at relative
momentum $q_{23}=0$  in the np QFS kinematic condition. As shown in the Fig.~\ref{Fig:wf} this corresponds to the point where the relative
wave function is sharply peaked in the case of S-wave, but zero in the case of P- or D- waves. 
At E$_{C}=897 ~{\rm MeV}$, as one moves away from the QFS point, that is, when $\theta_{C}$ increases from zero, the cross section starts to probe the wave function at larger relative momentum. For the S-wave case where the wave function is very narrow in momentum space, this means that the cross section will decrease very rapidly as one increases the core scattering angle,  as seen clearly in Fig.~\ref{Fig:ds-e69-ssca-spd}. For the case of a P-wave, it starts to probe nonvanishing values 
of the wave function reaching steeply its peak. Likewise for the D-wave as one moves away from the QFS point it starts to probe nonvanishing values of the wave function reaching less steeply its peak as also shown in Fig.~\ref{Fig:ds-e69-ssca-spd}. 

On the other hand the cross section calculated using SSA from the core probes the bound state wave function at relative momentum different from zero. This means that the contribution from the scattering from the core is small for the S-wave. As shown in the graph at this energy the scattering from the core gives an important contribution to the single scattering term in the case where the struck particle is bound to the core in P- or D-waves.
 
We would like to point out that the SSA  from the core is neglected in standard DWIA approaches
which are therefore inadequate for nonzero relative angular momentum wave functions of the bound pair. 

\begin{figure}
{\par\centering \resizebox*{0.35\textwidth}{!}
{\includegraphics{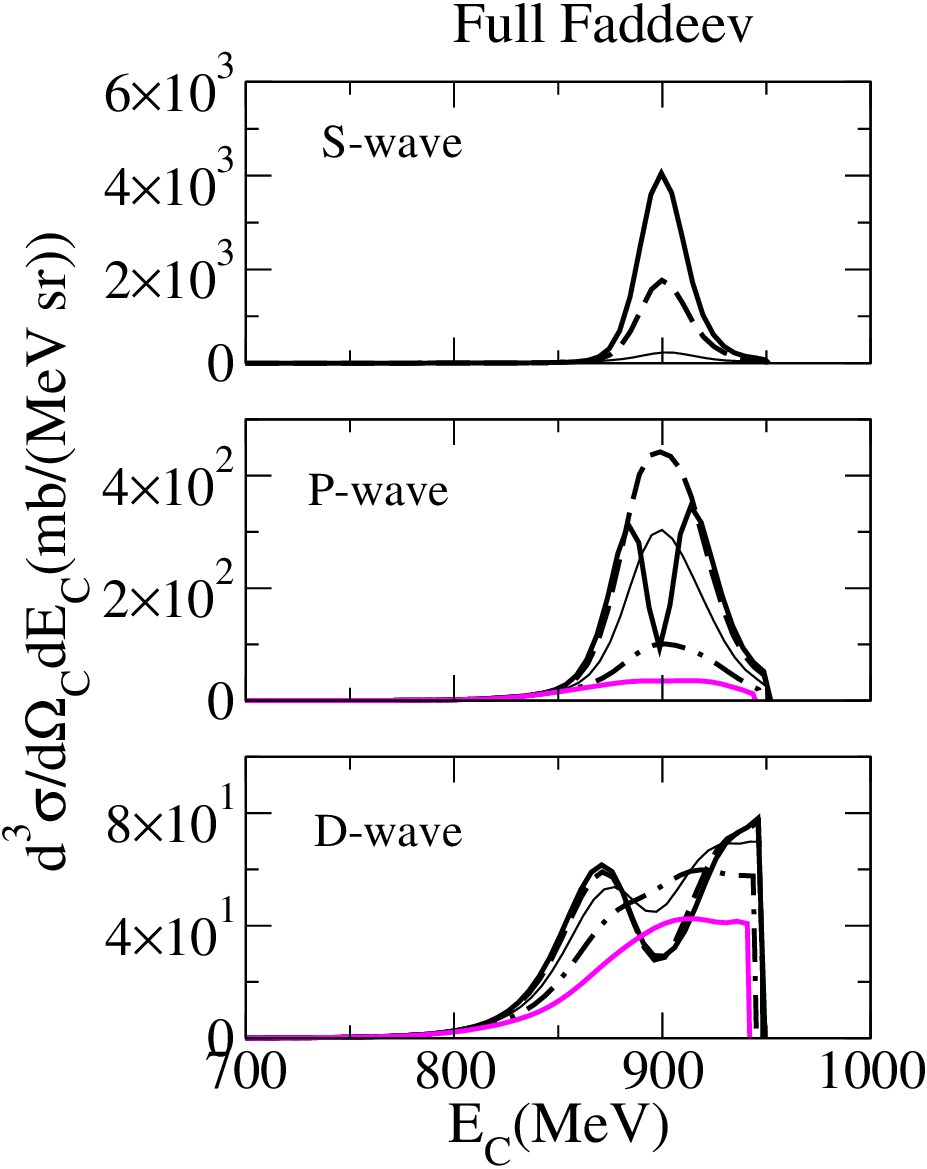}} \par}
\caption{\label{Fig:ds-e69-full}
 (Color online) Semi-inclusive cross section for the breakup
  p($^{14}$Be,$^{13}$Be)np at 69~MeV/u 
using the full multiple scattering calculations
for $\theta_{C}=0^\circ$ (dark thick solid line), 
$\theta_{C}=0.4^\circ$ (dashed line),
$\theta_{C}=0.8^\circ$ (dark thin solid line),
$\theta_{C}=1.2^\circ$ (dashed dotted line), and
$\theta_{C}=1.4^\circ$ (light solid line). 
The n-$^{13}$Be interaction is restricted to the $0^+$ partial wave.
}
\end{figure}

In Fig.~\ref{Fig:ds-e69-full} we show the semi-inclusive cross section for the breakup
p($^{14}$Be,$^{13}$Be)np at 69~MeV/u using full multiple scattering Faddeev/AGS 
calculations. The most striking feature from the graphs is the relative order of magnitude of the calculated observables for each single-particle configuration which is carried over from the SSA calculations shown in Fig.~\ref{Fig:ds-e69-ssca-spd}. The breakup observable calculated with the S-wave single particle state is one order (two-orders) of magnitude larger that the breakup observable calculated with the P-wave (D-wave). Therefore the magnitude of the semi-inclusive cross section  in quasifree conditions is a clear signature of the angular momentum of the valence nucleon. Comparing Figs.~\ref{Fig:ds-e69-ssca-spd}
and \ref{Fig:ds-e69-full}  we conclude that the SSA clearly overestimates the full multiple scattering results in all configurations at this energy regime. Therefore care should be taken into account when using truncated multiple scattering frameworks
such as DWIA as pointed out in Ref.~\cite{Crespo08}.

\begin{figure}
{\par\centering \resizebox*{0.35\textwidth}{!}
{\includegraphics{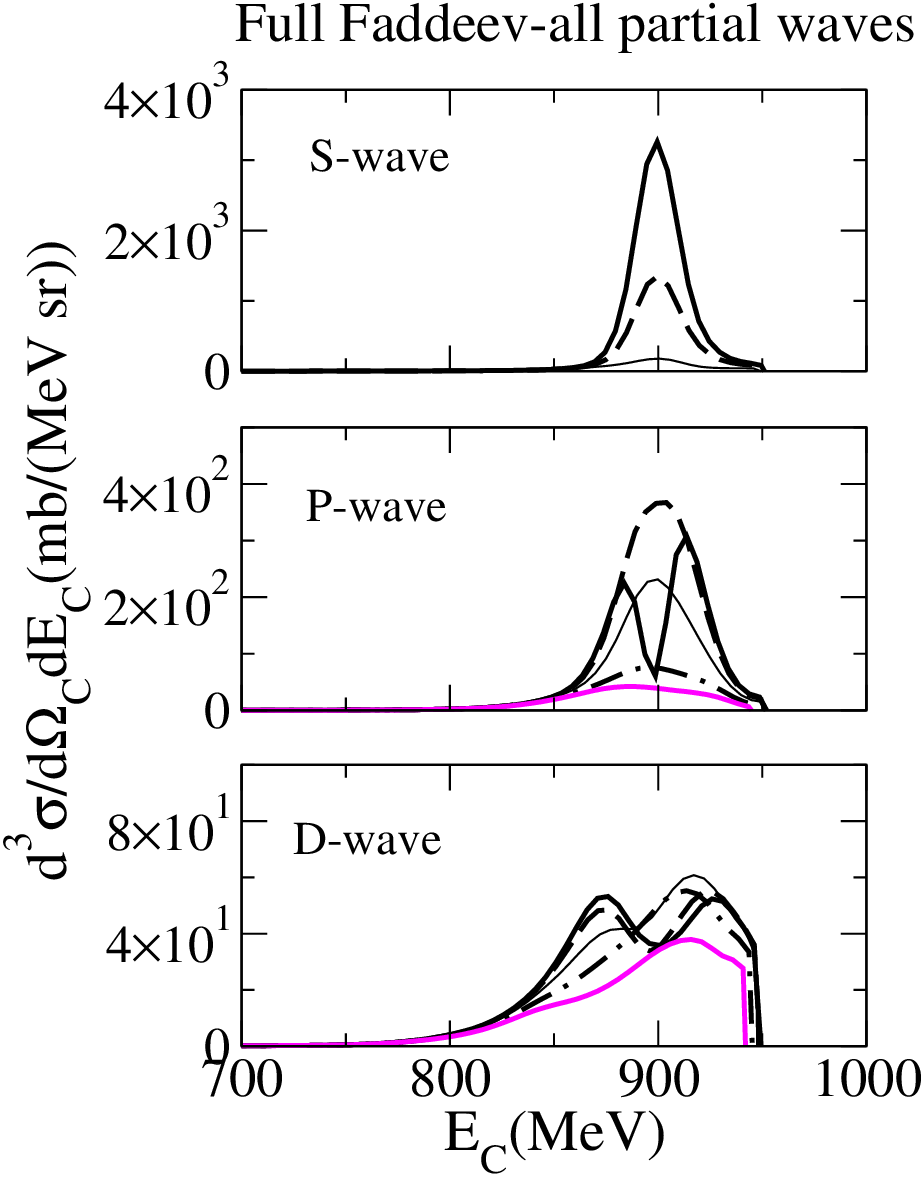}} \par}
\caption{\label{Fig:ds-e69-full-allpw}
 (Color online) Semi-inclusive cross section for the breakup
  p($^{14}$Be,$^{13}$Be)np at 69~MeV/u 
using the full multiple scattering calculations
for $\theta_{C}=0^\circ$ (dark thick solid line), 
$\theta_{C}=0.4^\circ$ (dashed line),
$\theta_{C}=0.8^\circ$ (dark thin solid line),
$\theta_{C}=1.2^\circ$ (dashed dotted line), and
$\theta_{C}=1.4^\circ$ (light solid line).
The n-$^{13}$Be interaction is included in all partial waves (see text).
}
\end{figure}

In order to assess the effect of introducing higher partial waves in the n-$^{13}$Be interaction we follow the prescription mentioned at the end of Section~\ref{The pair interactions} and use the potential corresponding to $V_c=32.922$ MeV in Table~\ref{Tab:n-13Be} in all partial waves other than $0^+$. For that matter we include n-$^{13}$Be relative angular momenta $\ell \le 3$. The results for the semi-inclusive cross section obtained from  the
 full multiple scattering Faddeev/AGS  calculation are shown in Fig.~\ref{Fig:ds-e69-full-allpw} and do not differ significantly from the results in Fig.~\ref{Fig:ds-e69-full}. This indicates that the effect of introducing the coupling to the other n-$^{13}$Be partial waves is small, and does not change the conclusions.

\begin{figure}
{\par\centering \resizebox*{0.40\textwidth}{!}
{\includegraphics{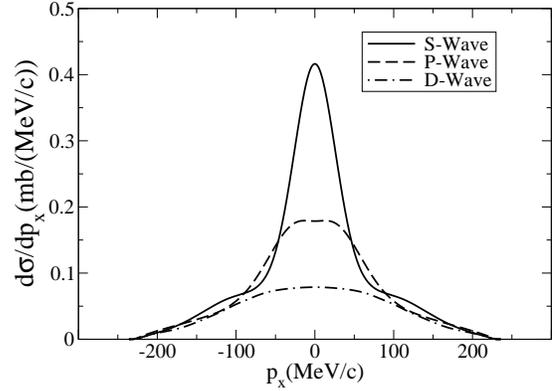}} \par}
\caption{\label{Fig:px-e69-full}
 (Color online) Transverse momentum distributions of
the $^{13}$Be = ($^{12}$Be+n) system in the p($^{14}$Be,$^{13}$Be)np reaction at 69 MeV/u. Results of the full Faddeev/AGS calculations with S-, P- and D-wave ($^{13}$Be+n) bound states are shown.
}
\end{figure}

In Fig.~\ref{Fig:px-e69-full} we show the inclusive transverse momentum distributions of
the $^{13}$Be = ($^{12}$Be+n) system in the $1/2^-$,  $1/2^+$, and $5/2^+$ states
at 69 MeV/u obtained from the full multiple scattering Faddeev/AGS calculations. These results show that the shape of the inclusive transverse momentum distribution does not provide a clear signature for the spin of the $^{13}$Be resonance, but instead, its magnitude at the peak does, as long as one has full control of the three-body dynamics.  Nevertheless the magnitude of the semi-inclusive cross section is far more sensitive to the orbital angular momentum of the ($^{13}$Be+n) $0^+$  bound state than the inclusive transverse momentum distribution of $^{13}$Be.

%%%%%%%%%%%%%%%%%%%%%%%%%%%%%%%%%%%%%%%%%%%%
%%%%%%%%%%%%%%%%%%%%%%%%%%%%%%%%%%%%%%%%%%%%

%%%%%%%%%%%%%%%%%%%%%%%%%%%%%%%%%%%%%%%%%%%%
%%%%%%%%%%%%%%%%%%%%%%%%%%%%%%%%%%%%%%%%%%%%

%%%%%%%%%%%%%%%%%%%%%%%%%%%%%%%%%%%%%%%%%%%%
%%%%%%%%%%%%%%%%%%%%%%%%%%%%%%%%%%%%%%%%%%%%

%%%%%%%%%%%%%%%%%%%%%%%%%%%%%%%%%%%%%%%%%%%%
%%%%%%%%%%%%%%%%%%%%%%%%%%%%%%%%%%%%%%%%%%%%

\section{Conclusions}

We have performed full Faddeev-type calculations  for 
one-neutron knockout reaction of $^{14}$Be on proton target
at 69 MeV/u incident energy. These results were compared with 
those corresponding to the single scattering approximation.
Inclusive transverse momentum distribution observables for the 
outgoing ($^{12}$Be+n) system were calculated. 
Semi-inclusive differential cross sections at the
np QFS kinematical conditions were also presented.

In this work we have considered the $0^+$ ground state of $^{14}$Be as a single particle state made up of ($^{13}$Be+n) bound in P-, S-, or D-wave depending on the $^{13}$Be  spin being $1/2^-$, $1/2^+$, or $5/2^+$, respectively.

We have found that the single scattering contribution from the core
is very significant in the case of P- or D-waves. In any case 
the single scattering  is a bad approximation
of the full results at this energy.
Thus, DWIA approaches which take into account the contribution between the
proton and the struck valence neutron, are inadequate in this case. Higher order multiple scattering contributions need to be taken into account as done in the full Faddeev/AGS approach.
The semi-inclusive breakup cross section resulting from the full Faddeev/AGS calculation with a 
 ($^{13}$Be+n) S-wave bound state is one order (two-orders)
of magnitude larger than in the case of the P-wave
(D-wave) bound state. Therefore  the magnitude of the semi-inclusive
cross section  in QFS conditions is a clear signature 
of the angular momentum of the valence nucleon if we can have a good control of the underlying three-body  dynamics.

\appendix
\renewcommand{\theequation}{\thesection.\arabic{equation}}
\setcounter{equation}{0}
\section{ }
 
In SSA, the initial relative momentum between particles 1 and 2
is
\be
{\bf q}_{12} = \frac{m_2}{M_{12}} \mbf{k}_{1} - 
\frac{m_1}{M_{12}}(- \mbf{k}_{1}- \mbf{k}^{\prime}_{3}).
\ee
In np QFS kinematical conditions
$\mbf{k}^{\prime}_3 = - \mbf{k}_1 \frac{m_3}{m_2+m_3}$ 
we get
\be 
[{\mbf{q}}_{12}]^{\rm QFS} = \frac{M m_2}{M_{12} M_{23}}\mbf{k}^{\prime}_{1} ~.
\ee
The two-body energy $\omega_{12}$, in the limit of zero binding energy for the pair (23), is given by
\be
\omega_{12} &=& E - \frac{{{k}^{\prime}_3}^2}{2 \mu_{3(12)}}
\nonumber \\
&=& \frac{{{k}_1^2}}{2 \mu_{1(23)}} -  
\frac{{{k}^{\prime}_3}^2}{2 \mu_{3(12)}} ~.
\ee
Under the QFS condition
\be
[\omega_{12}]^{\rm QFS} &=& \frac{{{k}_1^2}}{2 \mu_{1(23)}}
- \frac{{{k}_1^2}}{2 \mu_{3(12)}}  \frac{m_3^2}{{M_{23}}^2} 
\nonumber \\
 &=& {{k}_1^2} 
\frac{M^2 m_2}{2m_1 {M_{23}}^2 M_{12} } 
\nonumber \\
&=& E \frac{M m_2}{M_{23} M_{12} }~~,
\ee
leading to
\be
[\omega_{12}]^{\rm QFS}= \frac{\left[ {q^{2}_{12}}\right]^{\rm QFS}}
{2 \mu_{(12)}}~~.
\ee
Thus, in the single scattering term the matrix elements of the transition 
operator $t_{12}(\omega_{12})$
 are on the energy shell in the limit of zero binding for pair (23).

\vspace{0.5cm}
\section{ }

Let us consider the breakup reaction p((nC),C)np.
In this section we give the formulae for the momentum distributions
of the detected core C. 
The kinematics of particle C can be defined in terms of its momentum
$(K_x,K_y,K_z)$ or, alternatively, by its LAB energy and angular variables
$(E,\Omega)$ where, in this section, we drop the 
index of the core for simplification.
The semi-inclusive cross section is given by 
\be
\frac{d^3 \sigma}{d^3 K}= \frac{1}{m K} \frac{d^3 \sigma}{dE d\Omega}~,
\ee
with $m$ the mass of the core.
The Cartesian components of the momentum of the particle
can be expressed in terms of its spherical components ($K_x$,$K_y$,$K_z$)= 
($K \sin\theta \cos\phi$, $K \sin \theta \sin\phi$, $K \cos\theta$).
Alternatively one may define the set of cylindrical  momentum coordinates
$(K_\rho,\phi,K_z)=(\sqrt{K_x^2+K_y^2}, \tan^{-1}(K_y/K_x),K_z)$. One can write then
\be
d^3 K = dK_x dK_y dK_z = d^2K_{\rho} dK_z = K_{\rho} dK_{\rho} d\phi dK_z
\ee
Thus 
\be
\frac{d^2 \sigma}{d^2 K_{\rho}}=\int_{-\infty}^{+\infty}\frac{d^3 \sigma}{d^3 K}dK_z
\ee
From this double cross section we can calculate the inclusive perpendicular
momentum distribution
\be
\frac{d \sigma}{d K_{\rho}} = 
2 \pi K_{\rho}\int_{-\infty}^{+\infty}\frac{d^3 \sigma}{d^3 K}dK_z 
\ee
and the inclusive transverse momentum distribution
\be
\frac{d \sigma}{d K_x} = 2\int_{0}^{+\infty}\frac{d^2 \sigma}{d^2 K_{\rho}}dK_y
\ee

{\bf Acknowledgements:}
The authors would like to thank J. Tostevin for useful discussions.
The work of  A.D. is supported by the Funda\c c\~ao para a Ci\^encia e 
Tecnologia (FCT) grant SFRH/BPD/34628/2007 and all other authors by the
 FCT grant POCTI/ISFL/2/275.

\end{document}